\documentclass[preprint,12pt]{elsarticle}

\usepackage{amssymb}
\usepackage{subfigure}
\usepackage{amsmath}
\usepackage{framed}
\usepackage{booktabs}
\usepackage{color,soul}
\usepackage{algorithm}
\usepackage{multirow}
\usepackage{hyperref}
\usepackage{float}
\usepackage{algpseudocode}
\usepackage{comment}

\usepackage{ragged2e}
\justifying

\newcounter{algsubstate}

\newenvironment{algsubstates}
  {\setcounter{algsubstate}{0}%
   \renewcommand{\State}{%
     \refstepcounter{algsubstate}%
     \Statex {\footnotesize\alph{algsubstate}:}\space}}
  {}

\begin{document}
\begin{frontmatter}



\title{Incorporating Zero-Knowledge Succinct Non-interactive Argument of Knowledge for Blockchain-based Identity Management with off-chain computations}


\author[inst1]{Pranay Kothari}

\affiliation[inst1]{organization={Department of Computer Science and Engineering, Netaji Subhas University of Technology},%Department and Organization
            addressline={Delhi-110078},
            country={India}}

\author[inst1]{Deepak Chopra}
\author[inst1]{Manjot Singh}
\author[inst1]{Shivam Bhardwaj}
\author[inst1]{Rudresh Dwivedi}

\begin{abstract}
\par In today's world, secure and efficient biometric authentication is of keen importance. Traditional authentication methods are no longer considered reliable due to their susceptibility to cyber-attacks. Biometric authentication, particularly fingerprint authentication, has emerged as a promising alternative, but it raises concerns about the storage and use of biometric data, as well as centralized storage, which could make it vulnerable to cyber-attacks. In this paper, a novel blockchain-based fingerprint authentication system is proposed that integrates zk-SNARKs, which are zero-knowledge proofs that enable secure and efficient authentication without revealing sensitive biometric information. A KNN-based approach on the FVC2002, FVC2004 and FVC2006 datasets is used to generate a cancelable template for secure, faster, and robust biometric registration and authentication which is stored using the Interplanetary File System. The proposed approach provides an average accuracy of 99.01\%, 98.97\% and 98.52\% over the FVC2002, FVC2004 and FVC2006 datasets respectively for fingerprint authentication. Incorporation of $zk-SNARK$ facilitates smaller proof size. Overall, the proposed method has the potential to provide a secure and efficient solution for blockchain-based identity management. 
\end{abstract}


% \begin{highlights}
% \end{highlights}


\end{frontmatter}


\section{Introduction}
\subsection{Background}
\par Biometric authentication is a technology that uses an individual's unique biological characteristics, such as fingerprints, iris patterns, or facial features, to verify their identity. Fingerprint authentication is one of the most widely used biometric authentication techniques due to its ease of use, accuracy, and non-intrusive nature. Fingerprint authentication has many applications, including access control to secure facilities, mobile devices, and financial transactions. Fingerprint authentication is also used in forensics to identify suspects and in border control to screen travellers. While fingerprint authentication is a convenient and secure authentication method, it also raises certain privacy and security concerns with present technology \cite{dwivedi2017coprime}. The following are some problems with current fingerprint-based identity management systems: storage in a central database; privacy concerns as the data is considered sensitive personal information; risk of data misuse or access; and lack of transparency in the verification process as the algorithms used to match fingerprints are proprietary and not subject to scrutiny.\par
Blockchain technology is gaining widespread adoption across various industries, including finance, supply chain management, and identity management. The latter, in particular, is a critical application of blockchain technology, as it enables individuals to control their personal information while providing a secure and immutable record of their identity. Security, privacy, and efficiency issues with identity verification processes across multiple domains, from financial services to healthcare and beyond, can be effectively resolved by integrating blockchain with biometric identity management.
Blockchain technology's integration with biometric identity management not only improves security and privacy but also completely alters how people and businesses handle identity verification in the age of digital communication.

To address the security and privacy issues, we have used Zero-Knowledge Succinct Non-interactive Argument of Knowledge (zk-SNARK) for authentication. zk-SNARKs are a type of proof that allows users to demonstrate knowledge of a secret without revealing the secret itself. zk-SNARK uses two main metrics (proof size and verifier time) for the performance evaluation. This technique has numerous applications, including blockchain-based identity management solutions. By using zk-SNARKs, identity management solutions can significantly reduce the amount of data that needs to be stored on the blockchain, as the proofs themselves are much smaller than the data they represent. This, in turn, can improve the scalability of the blockchain network and reduce transaction times. Therefore, incorporating zk-SNARK for blockchain-based identity management with off-chain computations is a promising solution for the scalability challenges and data privacy and security issues facing blockchain-based identity management solutions. As blockchain technology continues to evolve, we will likely see more innovations in this area that will further improve the scalability and efficiency of blockchain-based identity management solutions.

\subsection{Motivation and Contribution}
\par This work introduces the zero-knowledge succinct non-interactive argument of knowledge to a blockchain-based identity management system. It utilizes a KNN-based approach to generate a cancelable template for fingerprint authentication, stored on InterPlanetary File System (IPFS). Incorporating zero-knowledge proofs into a blockchain-based biometric authentication system can enhance the security and trustworthiness of the system. By providing mathematical proof of the validity of biometric data, zero-knowledge proofs can prevent fraud and identity theft, making the system reliable and trustworthy. The primary objective is to aid the general public's identity management system, such as the national identity cards, as well as smaller-scale identity management systems used by schools, universities, and businesses. The key contributions of this work are as follows:
\begin{itemize}
    \item In this work, firstly we have used an efficient KNN-S-based algorithm for cancelable template generation.
    \item Additionally, we have incorporated zero-knowledge proofs for authentication for enhanced data security and privacy for biometric data storage.
    \item Finally, we have also integrated the InterPlanetary File Systems (IPFS) protocol in our authentication pipeline for storing our template in a decentralized and distributed file system.
\end{itemize}

\section{Related Work}
%\subsection{Blockchain-Based Identity Management Systems}
In recent years, there has been significant interest in developing blockchain-based identity management systems that provide secure, decentralized and tamper-proof verification of identity. One of the major challenges in developing such systems is ensuring privacy while maintaining the integrity of the data. To address this challenge, researchers have proposed incorporating zero-knowledge proofs (zk-SNARKs) \cite{blum1991noninteractive} and off-chain computations in blockchain-based identity management systems. 

To date, there have been several blockchain-based identity management systems \cite{guo2022zksnark}, \cite{loung2023privacy}, \cite{elgayyar2020federated}, \cite{xu2020mobile}, \cite{alsyed2019dns} and \cite{Mao2023} that have focused on safeguarding users’ privacy and ensuring anonymous interactions. ElGayyar et al. \cite{elgayyar2020federated} proposed a resilient and automatic blockchain-based federated identity management (IdM) system, where identities for users are automatically generated and audited by smart contracts. Users can control their identities and store them on the blockchain. However, this system's anonymity is entirely dependent on the anonymity of the blockchain. Additionally, because this system uses smart contracts to encrypt and decrypt user data or identities, user identities may be visible to the public due to the transparent nature of public blockchains. Xu et al. \cite{xu2020mobile} proposed a blockchain-based identity management and authentication scheme for mobile networks. The scheme is designed to be secure, efficient, and privacy-preserving. Kassem et al. \cite{alsyed2019dns} proposed a system, called DNS-IdM, designed to manage the digital identities of users and ensure secure communication between them while maintaining data privacy. The authors argue that traditional identity management systems have limitations in terms of security, privacy, and scalability, and therefore, a blockchain-based system can provide a more reliable and efficient solution. The paper provides a detailed description of the DNS-IdM architecture, including the roles of different components and their interactions. Guo et al. \cite{guo2022zksnark} proposed a novel biometric identification scheme based on the zero-knowledge succinct non-interactive argument of knowledge (ZK-SNARK). The scheme is designed to be efficient and secure, and it can be used to identify users based on their biometric data, such as fingerprints or iris scans.

Luong et al. \cite{loung2023privacy} proposed a privacy-preserving identity management system on the blockchain that enables users to handle their identity attributes while ensuring that their true identities remain concealed from all entities, including the identity provider. The game-based proof scheme is used to analyze and prove the system’s security requirements, by combining zk-SNARK, Shamir’s secret sharing (SSS), and several other cryptographic techniques. Mao et al. \cite{Mao2023} proposed a novel user-centric biometric authentication scheme called BAZKP. In this scheme, all the biometric data remain encrypted during the authentication phase, so the server never sees them directly. The server can still determine whether the Euclidean distance of two secret vectors is within a pre-defined threshold by calculation. The authors also implemented a privacy-preserving biometric authentication system based on BAZKP, and their evaluation demonstrates that it provides reliable and secure authentication.

%\subsection{Fingerprint Template Generation Techniques}
%Effland et al. \cite{tom2014local}, proposed a new method that is both efficient and robust over existing fingerprint hashing methods. The method partitions the fingerprint image into local subsets and extracts features from each subset. These features are then combined to form a robust and secure fingerprint hash. The proposed method is evaluated on standard fingerprint databases and outperforms existing methods in terms of recognition accuracy and computational complexity. The authors in \cite{sandhya2015} proposed a new method for creating cancelable fingerprint templates which are based on the k-nearest neighbours (k-NN) structures and do not require alignment of the fingerprint templates for matching. The authors argue that existing methods for protecting fingerprint templates are either too computationally expensive or have low recognition accuracy. The proposed method is evaluated on standard fingerprint databases and is shown to be more efficient and accurate than existing methods. The authors conclude that the proposed method can be used to effectively protect fingerprint templates in biometric systems.

\section{Proposed Approach}
The proposed work aims to develop a Blockchain-based identity management system which makes use of zk-SNARK for authentication and the previously proposed approach of using a KNN-based approach in \cite{sandhya2015} for fingerprint template generation.

\subsection{Fingerprint image Enhancement}
\begin{figure}[htbp]
        \centering
        \subfigure[]{\includegraphics[width=0.24\textwidth]{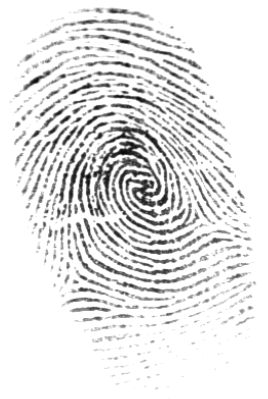}}
        \subfigure[]{\includegraphics[width=0.25\textwidth]{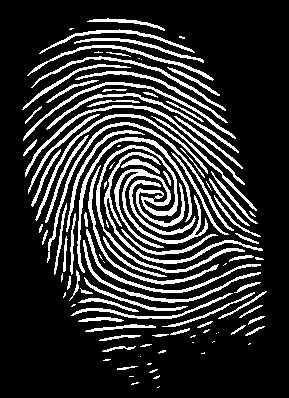}}
        \subfigure[]{\includegraphics[width=0.24\textwidth]{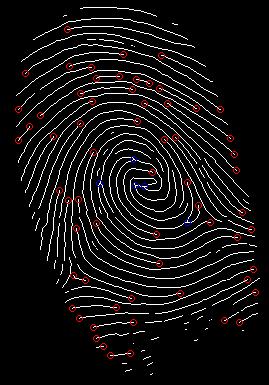}}
        \caption{(a) Original Fingerprint Image (b) Enhanced Fingerprint Image (c) Extracted minutiae points}
        \label{f1}
    \end{figure}

     Image enhancement is used to improve the quality of fingerprint images by enhancing the contrast, brightness and details of fingerprint images. This includes several steps such as image pre-processing, enhancement, and post-processing. During preprocessing, we filter the image to remove noise and separate the fingerprint region from the background. The ridges and valleys of the fingerprint are emphasized using enhancement techniques such as histogram equalization \cite{dorothyimgenhance2015}, binarization \cite{Sasirekha2015ACA}, contrast stretching \cite{firdausy2007}, and spatial filtering \cite{chavali2016}. Finally, we adjust the enhanced image to remove the distortion introduced during enhancement. %Fingerprint enhancement is important in fingerprint recognition systems to improve the accuracy of matching algorithms. 
     Figures \ref{f1} (a) and \ref{f1} (b) shows a fingerprint image before and after enhancement respectively.

\subsection{Minutiae points Extraction}
The minutiae features are widely used for fingerprint recognition, where the minutiae are local ridge characteristics, such as ridge endings and bifurcations. %The minutiae extraction \cite{Bansal2011MinutiaeEF} process involves analyzing the fingerprint image using various techniques, including image processing and computer vision algorithms, to extract the minutiae features. 
The algorithm involving crossing number computation \cite{virdaus2017} are used for minutiae extraction in our work. The feature vector obtained from the minutiae points includes the location, orientation, and type of minutiae points, which can be used to compare two fingerprints for similarity or dissimilarity. To ensure accurate recognition, a spurious minutiae removal process \cite{rachna2014} is also implemented to deal with the noise present in the input fingerprint image. Figure \ref{f1} (c) shows the extracted minutiae points in the enhanced fingerprint image.
\begin{figure}[!htbp]
    \centering
    \includegraphics[scale=0.27]{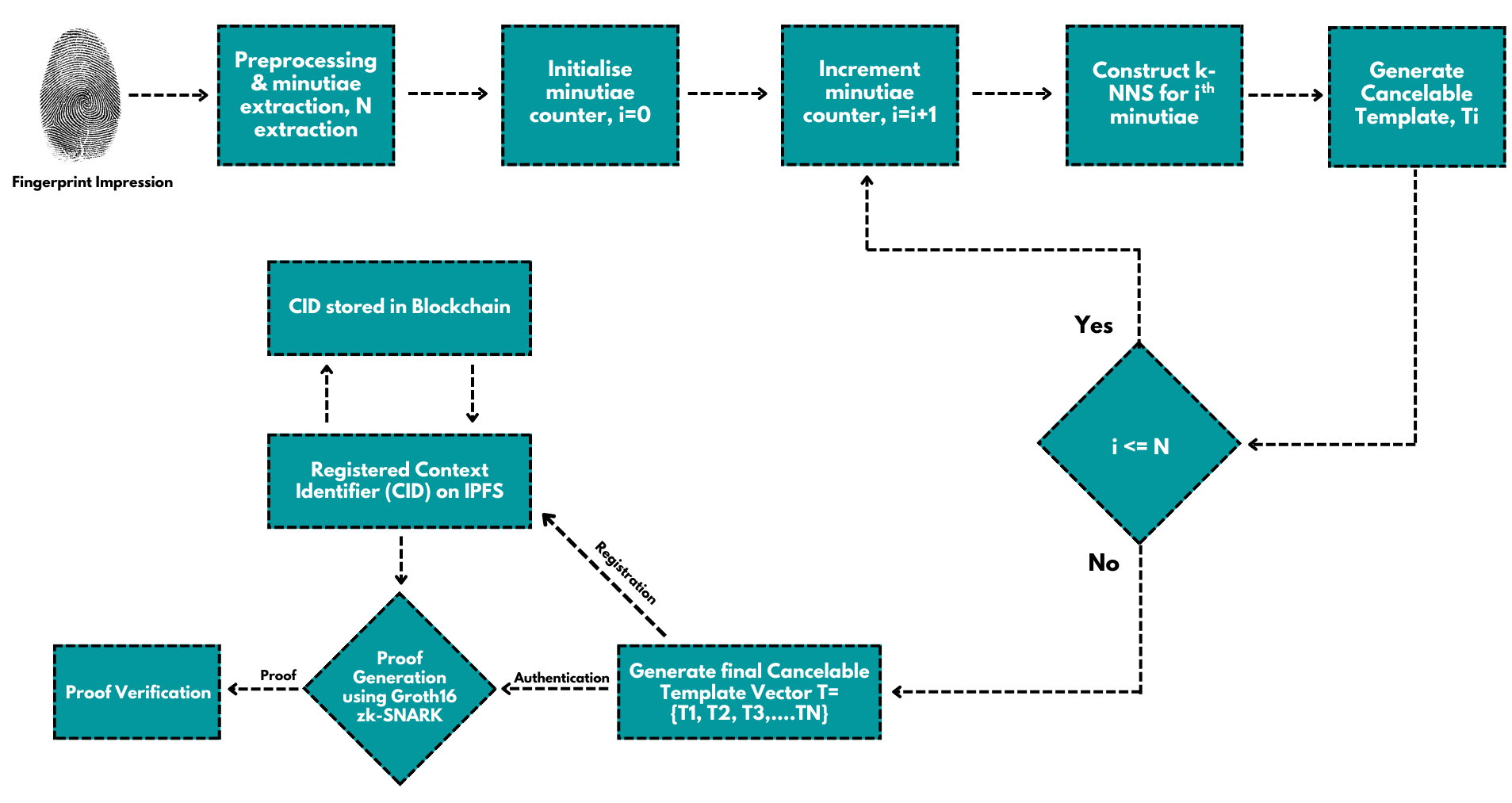}
    \caption{Flow diagram of K-NNS approach}
    \label{f3}
\end{figure}
\subsection{Generating K-NNS Based template}
This section discusses the process of creating a KNN structure for each minutiae point of the fingerprint as discussed in \cite{sandhya2015}. These structures are then utilized to facilitate user registration and authentication.

\subsubsection{Creating the k-Nearest Neighborhood structure}

Let,  
\begin{equation}
M_i=\left(x_i, y_i, \theta_i\right)_1^N
\end{equation}
represent the minutiae points that were extracted from the fingerprint image, where $N$ is the total number of minutiae in the fingerprint image, (x$_i$, y$_i$) are the $x$ and $y$ coordinates of minutiae point, and $\theta_i$ is the orientation of the minutiae point. 
As seen in Figure \ref{f3}, each minutiae point corresponding to index $i$ is taken into account as a reference minutiae $r$, and the (K-NNS$_r$) is calculated using the following equations:

\begin{equation}\label{eq2}
\chi=\left(x_j-x_r\right) \cos \theta_r+\left(y_j-y_r\right) \sin \theta_r
\end{equation}

\begin{equation}\label{eq3}
\gamma=\left(x_j-x_r\right) \sin \theta_r-\left(y_j-y_r\right) \cos \theta_r
\end{equation}

\begin{equation}\label{eq4}
d_{j r}=\sqrt{\chi^2+\gamma^2}, j=1 . . k
\end{equation}

\begin{equation}\label{eq5}
\theta_{j r}=\tan ^{-1}\left(\frac{\sin \theta_j+\sin \theta_r}{\cos \theta_j+\cos \theta_r}\right), j=1 \ldots k
\end{equation}

For a reference minutiae point $r$, the $d_{jr}$ (rotation-invariant distance) and the $\theta_{jr}$ (rotation-invariant average of the orientation) are calculated using Eq.(\ref{eq4}) and Eq.(\ref{eq5}) to form K-NNS$_r$ that can be represented by Eq. (\ref{eq6}).

    \begin{equation}\label{eq6}
k-N N S_r=\left(d_{11}, \theta_{11}\right),\left(d_{21}, \theta_{21}\right) \ldots . .\left(d_{k 1}, \theta_{k 1}\right)
\end{equation}
    
    where d$_{jr}$ is the rotation-invariant distance between the $j^{th}$ nearest neighbour minutiae point and reference minutiae point $r$ and $\theta_{jr}$ is the average orientation between $j^{th}$ minutiae point and reference minutiae point r.
    For each point minutiae point, we generate the same structure and Eq.(\ref{eq7}) may be used to express the final K-NNS vector generated for the $N$ points in the minutiae set.
\begin{equation}\label{eq7}
\begin{aligned}
k-N N S= & \left(k-N N S_1, k-N N S_2,\right. \\
& \left.\ldots k-N N S_N\right), i=1 \ldots N
\end{aligned}
\end{equation}
 
\subsubsection{Quantization of k-Nearest Neighborhood structure}

The calculated $kNNS_r$ is assigned to a two-dimensional array of specific size $W_x$ x $W_y$.The size of this matrix is calculated as $W_x$ = $\lfloor max(d_{jr})/C_x \rfloor$, 
$W_y$ = $\lfloor 2\pi/C_y \rfloor$.
The two-dimensional array is partitioned into sections that have specific sizes $C_x$, $C_y$.
Here, $W_x$ and $W_y$ denote the 2D array's cell counts, where $\lfloor . \rfloor$ stands for the floor function.

\subsubsection{Converting the 2D array into 1D bit string}

We fill up the ($d_{jr}$, $\theta_{jr}$) from the $kNNS_r$  onto the 2D array by calculating the corresponding indices ($x_i$, $y_i$) of the cell that includes the point ($d_{jr}$, $\theta_{jr}$), using Eq. (\ref{eq8}).

\begin{equation}\label{eq8}
\left\{\begin{array}{l}
x_i \\
y_i
\end{array}\right\}=\left\{\begin{array}{l}
\left\lfloor d_{j r} / C_x\right\rfloor \\
\left\lfloor\theta_{j r} / C_y\right\rfloor
\end{array}\right\}
\end{equation}
\\
where $x_i$, $y_i$ represent indices of the two-dimensional array. A cell's value is set to 1 if one point or more falls inside it; otherwise, it is set to 0. Sequentially reading this matrix forms a 1-D bit string ($B_t$) and this bit string's length is fixed and corresponds to the dimensions of the two-dimensional array. i.e., $t$ = $W_x$ × $W_y$

\subsubsection{Transforming 1D bit string to final template}

The bit string {$B_t$} cannot be stored directly because, if {$B_t$} is exposed to an attacker the $k-NNS_r$ will be disclosed. Hence we need to perform some transformation on $B_t$ before storing it.
At first, we apply Discrete Fourier Transform (DFT) on {$B_t$} to generate $V$, a complex vector. We use Eq. (\ref{eq9}) to perform a t-point DFT, as the size of $B_t$ is $t$.

\begin{equation}\label{eq9}
V=\sum_{s=0}^{t-1} B_t e^{-j 2 \pi i s / t}, i=0,1, \ldots . t-1
\end{equation}

The size of computed vector $V$ is $t$ × $1$. We need to protect $V$ as it is invertible. Hence, we apply a non-invertible transformation to $V$. Using a specific key unique to the user, a user-specific random matrix ($R$) is produced. The dimensions of this random matrix $R$ should be $p$ × $q$, where $q$ needs to be equal to $t$. Next, the multiplication of $V$ by $R$ results in a changeable template $T$ of dimensions $p$ × $1$ as follows:

\begin{equation}\label{eq10}
T=R \times V
\end{equation}

\subsubsection{Matching}
For Authentication, we compare the stored fingerprint template and query fingerprint template, generating a score ranging from $0$ to $1$. This comparison is done in two steps.
We first calculate a local matching score between the templates. Then using these local matching score s we compute the final global matching score that ranges from $0$ to $1$. $1$ indicates a perfect match and $0$ indicates a total mismatch.

\begin{enumerate}
\item \textbf{Local matching score:}
We calculate distance between $T_i$ and $\vartheta_j$ by Eq. (\ref{eq11})
\begin{equation}\label{eq11}
{d\left(T_i, \vartheta_j\right)=} \frac{\left\|T_i-\vartheta_j\right\|_2}{\left\|T_i\right\|_2+\left\|\vartheta_j\right\|_2}
\end{equation}
Using this distance $d\left(T_i, \vartheta_j\right)$, the generated local matching score between the query and stored fingerprint template is given by equation \ref{eq12}-
\begin{equation}\label{eq12}
L M S\left(T_i, \vartheta_j\right)=1-d\left(T_i, \vartheta_j\right)
\end{equation}
The query and enrolled templates both have multiple entries in them. Let the enrolled fingerprint template
and the query fingerprint template be $T$ = $(T_1, T_2...T_e)$ and $\vartheta = (\vartheta_1, \vartheta_2, ...\vartheta_q)$ respectively. 
where $e$ and $q$ are the numbers of entries (number of minutiae points) in enrolled and query fingerprint templates respectively. We calculate the local matching score (LMS) between individual entries of the enrolled and query fingerprint template.
\item \textbf{Global matching score:}
By calculating LMS as shown in Figure \ref{f3} we matched every $T_i$ in $T$ = $(T_1, T_2...T_e)$ with every $\vartheta_j$ in $\vartheta = (\vartheta_1, \vartheta_2, ...\vartheta_q)$ which generates a similarity matrix $S(Ti,\vartheta_j)$ of dimensions $e * q$.
We calculate the global matching score (GMS) from this similarity matrix by using equation (\ref{eq13}). 
\begin{equation}\label{eq13}
G M S=\frac{\sum_{i=1}^e \sum_{j=1}^q S\left(T_i, \vartheta_j\right)}{\eta S\left(T_i, \vartheta_j\right)}
\end{equation}
where the number of non-zero elements in similarity matrix S($T_i$, $\vartheta_j$) is denoted by $\eta$.
\end{enumerate}

\subsection{Storage of Template using Interplanetary File System}
% Once we generate a final cancelable K-NNS template, we store the template on Moralis, which uses the IPFS protocol. IPFS, or InterPlanetary File System, is a decentralized file system that allows computers to store and access data on a peer-to-peer network. This implies that the data is not governed by any central authority, which makes it further secure and reliable than traditional centralized file systems. \par
% After storing, we receive a Context Identifier (CID) which could be used to access our KNNS template on IPFS. The CID is further stored on the Blockchain using smart contracts, deployed on Goerli Ethereum Testnet and in return, we get a unique User Key.
Once we generate a final cancelable K-NNS template, we store the template on the IPFS protocol. IPFS, or InterPlanetary File System, is a decentralized file system that allows computers to store and access data on a peer-to-peer network. This signifies that the data is not stored on any single server, but rather distributed across a network of computers. This makes IPFS more secure and reliable than traditional centralized file systems, as it is not vulnerable to single points of failure.

To store the template on IPFS, we first need to create a CID or Context Identifier. A CID is a unique identifier for a piece of data on IPFS. Once we have created a CID, we can then store the template on IPFS by using the CID as a key. 
After storing the template on IPFS, we can then store the CID on the blockchain. The blockchain is a distributed ledger that records transactions across a network of computers. By storing the CID on the blockchain, we ensure that the template is tamper-proof and secure. To store the CID on the blockchain, we utilize smart contracts. Smart contracts are self-executing contracts that are stored on the blockchain. On deploying a smart contract, we specify the CID as a data field. Once the smart contract is deployed, the CID will be stored on the blockchain. In return for storing the CID on the blockchain, we receive a unique user Key. \par
Table \ref{t1} provides the list of all the notations along with the description used in this paper.
\begin{table}[h]
\centering
\caption{Symbols and their descriptions}\label{t1}
\begin{tabular}{ll}
\toprule
\textbf{Notation} & \textbf{Description} \\
\midrule
$n$& Number of variables \\
$m$& Number of constraints \\
$p$& Public Inputs \\
$num$ & The number of points \\
$\mathbb{S}$ & The sum of local conditions \\
$\mathbb{S}'$ & The sum of the local conditions and the global condition \\
$t$ & The threshold value \\
$LSM$ & Matrix of Local Similarity Scores \\
$LC$ & Local Conditions \\
$\mathbb{Z}$ & A boolean signal indicating the satisfied global condition \\
$\mathbb{F}$ & Finite field \\
$\omega$ & Witness \\
$A, B \text{ and } C$ & Matrices in R1CS \\
$G_1, G_2 \text{ and } G_3$ & Layers of Groth16 \\
$\tau$ & Security parameter of the proof system \\
$\alpha$ & Modulus of the proof system \\
$\beta$ & Error probability of the proof system \\
$\delta$ & Soundness error \\
$\gamma$ & Witness size to generate a proof \\
\bottomrule
\end{tabular}
\end{table}

\subsection{Proof Generation and Verification using zk-SNARK}
SNARK stands for Succinct Non-Interactive Arguments of Knowledge and it is a succinct proof to prove that a particular statement is true.
For example, a prover might claim that he knows a particular message m such that SHA256($m$) is equal to 0 and wants to prove to others that he knows such a message $‘m’$. \par
% Let’s say ‘m’ is a gigabyte-long message, the trivial proof that the prover knows ‘m’ is to send the message to the verifiers and then the verifiers can verify that SHA256(m) is equal to 0.
% The problem with this trivial proof is that it is not short, it would be a gigabyte long if the message happens to be a gigabyte long. It’s also not fast to verify since the verifier has to hash the entire message to verify that it’s equal to zero.
% Our goal is to build a proof that is going to be very short, that is a few kilobytes and it’s going to be superfast to verify, that is a few milliseconds and the verifier will be convinced that the prover knows such a message ‘m’. SNARK allows the prover to generate such proof.
\par There is an extension to SNARK called the Zero-Knowledge SNARK where not only the prover wants to convince the verifier that he knows a message $‘m’$ that satisfies the property, but the prover also wants to do it without revealing anything about the message $‘m’$ other than the fact that it’s SHA256 hash is equal to $0$.
It allows proving that certain statements are true without revealing anything about why the statements are true. The steps followed in zk-SNARK are discussed in the following subsections.

\subsubsection{Generating Arithmetic Circuit}
 \par zkSNARK's allow for the proof of computational claims, but they cannot be immediately applied to the computational problem; the statement must first be translated into the correct form. Specifically, zkSNARK demands that the computational statement be represented by an arithmetic circuit.
 
An arithmetic circuit is a type of circuit which is used to verify the validity of a computation. It consists of a collection of gates, with wires connecting the gates that perform some arithmetic operations such as addition, subtraction and multiplication. The gates are connected in a particular way to ensure that the computation is performed correctly. An arithmetic circuit
takes some input signals, performs computation and gives out output signals.
The output of every gate is considered an “intermediate signal” if it is not an output signal.
The arithmetic circuit defines a multivariate polynomial, a polynomial in the input variables and gives a recipe for evaluating the polynomial. \par

In our work, we have designed two arithmetic circuits $A$ and $B$. Circuit $A$ is a threshold check circuit. It takes as input a threshold value and a local similarity matrix, and outputs a boolean value indicating whether the global matching score is greater than the threshold. Circuit $B$ contains two templates: LessEqualThan and GreaterThan. These templates implement the less than or equal to and greater than comparators, respectively. The templates take as input two numbers, and output a boolean value indicating whether the first number is less than or equal to (or greater than) the second number. The algorithm of arithmetic circuit $A$ is shown in Algorithm \ref{al}.

\begin{algorithm}[!htb]
\caption{Arithmetic circuit of A}\label{al}
   \textbf{Private inputs:} $\mathbb{S}$, $num$, $\mathbb{S}'$  \\
   \textbf{Public inputs:} $t$, $LSM$  \\
   \textbf{Output:} $\mathbb{Z} = \mathbb{S} >=$ $t$ x $num$ \\
\begin{algorithmic}[1]
  \State Initialize $\mathbb{S}$ $\longleftarrow$ 0 and $k$ $\longleftarrow$ $0$
  \State For each entry in $LSM$: 
  \begin{algsubstates}
    \State $num$ $\longleftarrow$ $num + 1$ 
    \State Add entry to $\mathbb{S}$ 
    \State Create LessEqualThan component and set inputs to entry and 100 
    \State Store output of LessEqualThan component to $LC[k]$ 
    \State Add $LC[k]$ to $\mathbb{S}'$ 
\end{algsubstates}
  \State Set $LCS$ to $\mathbb{S}'$
  \State Initialize $t'$ $\longleftarrow$ $t x num$
  \State Create GreaterThan component and set inputs to $\mathbb{S}$ and $t'$
  \State Store output of GreaterThan component to $\mathbb{Z}$
\end{algorithmic} 
\end{algorithm}

% \begin{framed}
% function isGlobalConditionSatisfied(threshold, LocSimMatrix): \\
%   sum = 0 \\
%   num\_of\_pts = 0 \\
%   for i in range(len(LocSimMatrix)): \\
%     for j in range(len(LocSimMatrix[0])): \\
%       sum += LocSimMatrix[i][j] \\
%       num\_of\_pts += 1 \\
%   threshold\_x\_no\_of\_pts = threshold \times num\_of\_pts \\
%   isGlobalConditionSatisfied = sum \geq threshold\_x\_no\_of\_pts \\
%   return  isGlobalConditionSatisfied \\
% \end{framed}

% In our approach, the mathematical equation for the global condition is of the form
% \begin{equation}\label{eq14}
% \text{isGlobalConditionSatisfied} = \sum_{i=1}^{n} x_i \geq \text{threshold} \times \text{num\_of\_pts}
% \end{equation}

% This equation states that the global condition is satisfied if the sum of the local conditions is greater than or equal to the product of the threshold and the number of points.

\subsubsection{Converting the circuit to R1CS}
R1CS stands for Rank-1 Constraint System. It is a way of expressing constraints on a set of variables that satisfy a system of linear equations. R1CS can be thought of as a matrix representation of the arithmetic constraints, where the rows of the matrix represent each equation, and the columns represent the variables. R1CS stands for Rank-1 Constraint System. It is a way of expressing constraints on a set of variables that satisfy a system of linear equations, where each equation is of the form:

\begin{equation}\label{eq14}
  a \times b=c  
\end{equation}

R1CS can be thought of as a matrix representation of the arithmetic constraints, where the rows of the matrix represent each equation, and the columns represent the variables. R1CS is a popular choice for ZKPs because it can be efficiently verified using linear algebra techniques and has relatively tiny proof sizes. 

The arithmetic circuit is converted to R1CS. This involves creating constraints that enforce the computation described by the circuit. The R1CS should have three sets of variables: primary inputs, intermediate inputs, and outputs. The primary inputs correspond to the inputs to the circuit. The intermediate inputs correspond to the intermediate variables in the circuit, and the outputs correspond to the final value that satisfies the equation. 
For an addition and multiplication gated arithmetic circuit $c(x,y)$ over $\mathbb{F}$, where some input/output wires are designated to specify statement $x$ and the other wires to specify a $\omega$. 
The prover aims to persuade the verifier that there exists the $\omega$ such that

\begin{equation}\label{eq15}
  c(x, \omega)=y
\end{equation}

The statement x is true if the prover is aware of the $\omega$ and $z$ that fulfil the R1CS requirements. To create an R1CS instance A, B, and C are formulated in a way that allows for the existence of $z$ that satisfies the equation \ref{eq16} if and only if $c(x, \omega) = y$:

\begin{equation}\label{eq16}
  (z \cdot A) \circ(z \cdot B)=z \cdot C
\end{equation}

The sizes of A, B, and C in the R1CS instance are directly related to the number of gates present in the arithmetic circuit $c$.
\\The $z$ is assigned an m-length vector where
% The dimension of matrices A, B, and C in the R1CS instance is proportional to the number of gates in arithmetic circuit C. Ultimately, the R1CS instance will have n constraint equations. 
% We set the solution vector z to be an m-length vector, where 
$z_0$ = 1. Each remaining entry of $z$ represents either the $x$ or $\omega$.

\begin{equation}\label{eq17}
   x=\left\{z_1, \ldots . z_l\right\} \text { and } \omega=\left\{z_l+1, \ldots . z_m\right\} 
\end{equation}
where, $l \leq m$

We compute a triple (a,b,c) for each gate, where, $a=A_k$ for row k of A, as follows:

\begin{enumerate}
    \item { Input gate in C: For input $x_i$ we want the entry $z_k$ in our R1CS instance to assert that $z_k=x_i$. Thus, we set $a=A_k$ to be the basis vector $e_1=\mathbb{F}^l$, $B_k=e_k \in \mathbb{F}^l$, and $C_k=x_i.e_1$}
    \item {Multiplication operation logic gate in C: Let $j_1$,$j_2$ be the input nodes for the multiplication gate. We want entry $z_k$ to assert $Z_{j_1} \cdot Z_{j_2}=Z_k$. Thus, we set $A_k=e_{j_1}\in \mathbb{F}^{n-1}$,$B_k=e_{j_2}\in \mathbb{F}^{m-1}$, and $c_k=e_k$}
    \item {Additive operation logic gate in C: Let $j_1$,$j_2$ be the input nodes for the addition gate. We want entry $z_k$ to assert $Z_{j_1} + Z_{j_2}=Z_k$.Thus, we set $A_k=e_1\in \mathbb{F}^{n-1}$, $B_k=e_{j_1}+e_{j_2}\in \mathbb{F}^{m-1}$, and $c_k=e_k$}
\end{enumerate}

We do this until the full R1CS instance is created such that the solution vector z will satisfy equation (\ref{eq16}) if and only if z is correct. A circuit may have several gates, resulting in numerous constraints in an R1CS instance. Hence, we convert the R1CS instance to a QAP (Quadratic Arithmetic Program) to save computational overhead.

We fix a basis $x \in \mathbb{F}^m$ and for each constraint define polynomials such that $A_i(x_j) = {A_i}_j$, $B_i(x_j) = {B_i}_j$ and $C_i(x_j) = {C_i}_j$.

Now we have, \\
\begin{center}
	$A(X) = \displaystyle\sum\limits_{i=0}^n \omega_i \cdot A_i(X)$ \\
	$B(X) = \displaystyle\sum\limits_{i=0}^n \omega_i \cdot B_i(X)$ \\
	$C(X) = \displaystyle\sum\limits_{i=0}^n \omega_i \cdot C_i(X)$ \\
\end{center}

The constraints are now equivalent to $A(x_i) \cdot B(x_i) = C(x_i)$, where $A(x_i)$, $B(x_i)$ and $C(x_i)$ are polynomials of degree $m-1$, $m-1$ and $(m-1) \cdot (m-1)$ respectively. This form is called a Quadratic Arithmetic Program and it can be verified by the existence of $H(X)$ that satisfies

\begin{equation} \label{eq18}
    A(X) \cdot B(X) - C(X) = H(X) \cdot Z(X)
\end{equation}

Thus, $m$ sets of constraints are converted into a single constraint.
If there is a solution, then there will be a proper assignment of $S$ and that assignment of $S$ will be in such a way that equation \ref{eq18} holds and in case of an incorrect assignment, the equation \ref{eq18} will not hold.
% \begin{framed}
% function isGlobalConditionSatisfied(threshold, LocSimMatrix): \\
%   sum = 0 \\
%   num\_of\_pts = 0 \\
%   for i in range(len(LocSimMatrix)): \\
%     for j in range(len(LocSimMatrix[0])): \\
%       sum += LocSimMatrix[i][j] \\
%       num\_of\_pts += 1 \\
%   threshold\_x\_no\_of\_pts = threshold \times num\_of\_pts \\
%   isGlobalConditionSatisfied = sum \geq threshold\_x\_no\_of\_pts \\
%   return  isGlobalConditionSatisfied \\
% \end{framed}

\subsubsection{Generating a trusted setup}
The next step is to perform a trusted setup \cite{buterin2022}, which involves creating a set of public parameters that will be used to generate and verify the proof. This setup should be done in a way that ensures the parameters cannot be manipulated or compromised and the prover is not able to break the security of the system. 

There are various approaches to do a trusted setup, but one common approach is to employ a multi-party computation protocol with numerous players, each of whom adds randomness to the setup. Groth16 \cite{cryptoeprint:2016:260} requires a per-circuit trusted setup. The trusted setup consists of 2 parts.

\begin{itemize}
    \item The powers of tau, which is independent of the circuit
    \item Circuit-specific program
\end{itemize}

The main idea behind the Powers of Tau is that by having multiple participants contribute randomness to the generation of the parameters, it becomes much more difficult for any individual machine/node or group to tamper with them or compromise the entire security of the system. It is called `Powers of Tau' as we use the powers of the respective generator functions of the numbers within a group, belonging to each actor/participant in Multi-Party Computation, the final parameter set we are generating using this is denoted as Tau.

The primary job of Multiparty computation schemes is to check whether no single entity generates the entire Common Reference String (CRS), or in other words, can gain entire knowledge of the underlying math of the CRS. This is achieved by allowing more and more parties to participate in the MPC scheme, but in such a way that we only need a minority of the participants to be honest. The more honest participants we find, the more randomness is added to the CRS generation, and the more secure the system gets.

\subsubsection{Computing the proving key}
Using the R1CS and the trusted setup, the proving key is computed. The proving key is a set of parameters that will be used to generate the proof that can be verified by anyone who knows the verification key and is kept private to ensure the security of the proof.

\subsubsection{Generating witness}
A witness is a set of variable assignments that satisfy the constraints of the system being proven and produce the desired output. The witness is a secret known only to the prover, and the prover must use the witness to generate proof without revealing any information about the witness itself.
The prover must take steps to ensure that the witness is kept secret. If the witness is revealed, the security of the system can be completely undermined.

% The witness is stored in a file with the extension .wtns and is encoded in a binary format compatible with snarkjs, which is then used to create the proof.

\subsubsection{Generating the proof}
Using the witness, proving key, \& the R1CS, the proof is generated. This typically involves using a set of pairing operations and exponentiations to generate a set of elliptic curve points that demonstrate the knowledge of the witness without revealing any information about the witness itself.
Pairing is a bilinear operation between two elliptic curve groups. Specifically, it takes as input two points P and Q on two different elliptic curves and outputs a scalar value that represents the bilinear pairing between the two points.

% In Groth16, the pairing operation is used to combine elliptic curve points from different parts of the proof to form a single value that can be used to verify the correctness of the proof.

% Exponentiation is a basic arithmetic operation that involves raising a number to a certain power. In Groth16, exponentiation is typically performed on elliptic curve points and is used to perform operations such as multiplying or dividing the points by scalar values.

In the process of generating a Groth16 proof, exponentiation is used to compute intermediate values that are used to generate the final proof. Specifically, the witness values and other parameters from the public parameters are exponentiated to generate intermediate values, which are then combined using pairing operations to generate the final proof.

Given public input $\omega = [1, \omega_1, \omega_2, ... , \omega_n]$ that satisfies the constraints, degree $m$ polynomials $A, B, C, H$ are computed first as mentioned in equation \ref{eq18}. Then, private witness polynomials are computed as:

\begin{center}
$L(X) = \displaystyle\sum\limits_{i \in [p,m)} \omega_i \cdot L_i(X)$ 
\end{center}

\begin{center}
Generating two random values $u, v$, to compute $A, B$ and $C$: \\
$A = (\alpha + u \cdot \delta + A(\tau)) \cdot G_1$ \\
$B = (\beta + v \cdot \delta + B(\tau)) \cdot G_2$ \\ 
$C = \left(\begin{array}{cc}
	& \delta^-1 \cdot (L(\tau) + H(\tau) \cdot Z_x(\tau))   \\
	& + v \cdot (\alpha + u \cdot \delta + A(\tau)) \\
	& + u \cdot (\beta + v \cdot \delta + B(\tau)) \\ 
	& - u \cdot v \cdot \delta \\
\end{array}\right) \cdot G_1$ 
\end{center}

The proof is the triplet $(A, B, C)$ and it is malleable.

\subsubsection{Verifying the proof}
The verification of proof in Groth16 involves computing a set of checks on the proof \cite{cryptoeprint:2016:260} and the public parameters to ensure that the proof is valid. These checks involve computing certain exponentiation and pairings between elliptic curve points to ensure that the proof satisfies certain mathematical properties.

A set of checks are done on the public parameters to ensure that they are valid and correspond to a valid set of polynomials that represent the constraints of the computation. Then the set of intermediate values is computed based on the proof and the public parameters. Then a set of checks is computed on the intermediate values to ensure that they are valid. This involves checking that the intermediate values satisfy certain mathematical properties that are required for the proof to be valid. The results of the check on the intermediate values to the result that is included in the proof are compared. If the two results match, then the proof is considered valid. The verification process is significantly faster than the generation process, which is important for real-world applications where proofs may need to be verified quickly and efficiently.

The verification is done by the smart contract written in Solidity and each transaction is recorded on the blockchain.

Given $\pi = (A, B, C)$ and public inputs $\omega = [1, \omega_1, \omega_2, ... , \omega_n]$. Public inputs are aggregated as:

\begin{equation} \label{eq19}
    \overline{L} = \displaystyle\sum\limits_{i \in [p,m)} \omega_i \cdot (\gamma^{-1} \cdot L_i(\tau) \cdot G_1)
\end{equation}

then we check

\begin{equation} \label{eq20}
    e(A, B) = e(\alpha \cdot G_1, \beta \cdot G_2) + e(\overline{L}, \gamma \cdot G_2) + e(C, \delta \cdot G_2)
\end{equation}

Essentially, equation \ref{eq20} verifies $A(\tau) \cdot B(\tau) = C(\tau) + H(\tau) \cdot Z_x(\tau)$ with some additional parameters.

\subsection{Fingerprint Authentication }
\begin{figure}[htbp]
    \centering
    \includegraphics[scale=0.3]{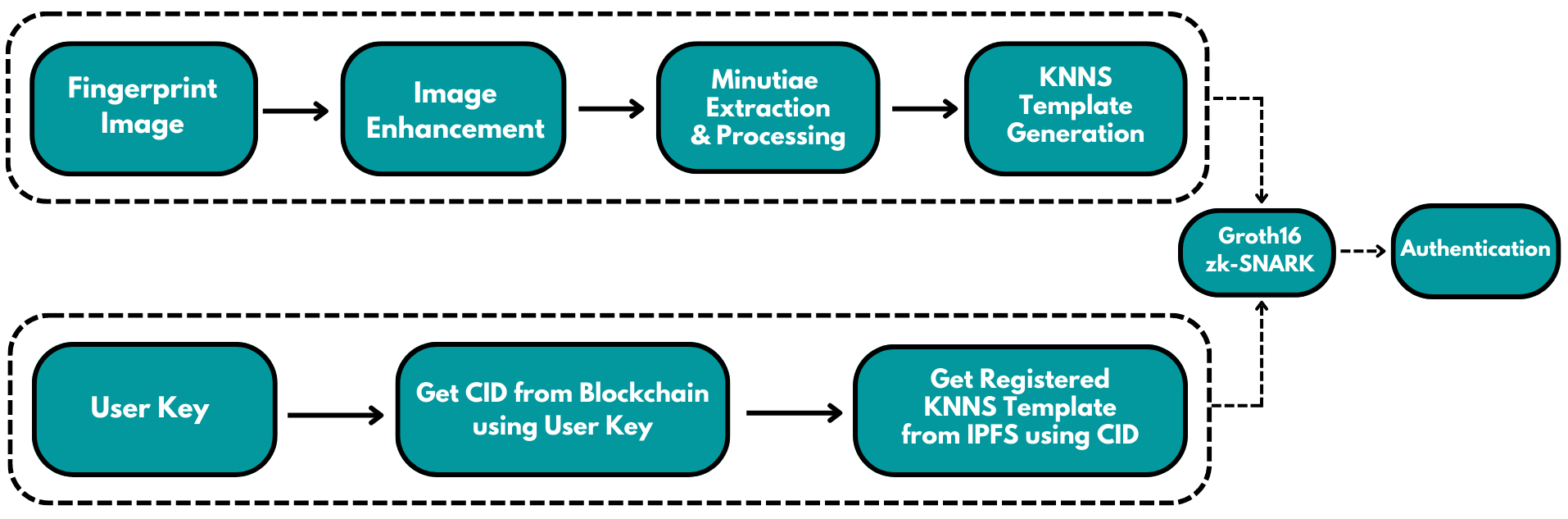}
    \caption{Flow diagram of Authentication}
    \label{f4}
\end{figure}

Two fingerprint templates are used in the authentication process of a user. The first template is known as the "Test User's Template" and is generated from the test user's fingerprint image. The second template which is the "Stored Template" is obtained using the user key and the CID retrieved from the blockchain from the already registered templates stored in IPFS. 
The Test User's Template is generated through a series of steps. The first step is the acquisition of the fingerprint from the user. The KNNS template is generated from this image. This template is used as the Test User's Template in the authentication process. The stored template, on the other hand, is obtained using the user key to retrieve the CID from the blockchain. The CID is then used to retrieve the registered KNNs template from IPFS. This template is used as the stored template in the authentication process. A Similarity matrix of the local matching score is generated between the stored and user template and a zero-knowledge proof is generated using the Groth16 algorithm to be sent to the verifier for proving that the local similarity matrix satisfies the necessary constraints and the global matching score is greater than the chosen threshold value. Figure \ref{f4} depicts the flowchart of the authentication process.

\section{Experimental Results and Analysis}
\subsection{Dataset}
    The FVC2002, FVC2004 and FVC2006 Fingerprint Datasets have been widely used as a reference database for fingerprint recognition.  The dataset has been established as an appropriate benchmark to assess the efficiency of fingerprint recognition algorithms by means of a range of sensors and acquisition conditions.
    Table \ref{t3}, \ref{t4} and \ref{t5} describes the composition of the FVC2002, FVC2004 and FVC2006 datasets respectively on the basis of sensor types, image dimensions, number of images per set, and the resolution.

    \begin{table}[ht!]
        \centering
            \caption{FVC2002 dataset}\label{t3}
            \scalebox{0.7}{
                \begin{tabular}{| c | c | c | c | c | c |}
                \hline
                \textbf{} & \textbf{Sensor Type} & \textbf{Image Dimensions} & \textbf{Set A (d x w)} & \textbf{Set B (d x w)} & \textbf{Resolution}\\ \hline
                \textbf{DB1} & Optical Sensor & 388x374 (142 Kpixels) & 8x100 & 8x10 & 500 dpi \\ \hline
                \textbf{DB2} & Optical Sensor & 296x560 (162 Kpixels) & 8x100 & 8x10 & 569 dpi \\\hline
                \textbf{DB3} & Capacitive Sensor & 300x300 (88 Kpixels) & 8x100 & 8x10 & 500 dpi \\\hline
                \textbf{DB4} & SFinGe v2.51 & 288x384 (108 Kpixels) & 8x100 & 8x10 & about 500 dpi \\ \hline
                \end{tabular}
            }
        \end{table}

        \begin{table}[ht!]
        \centering
            \caption{FVC2004 dataset}\label{t4}
            \resizebox{\textwidth}{!}{
                \begin{tabular}{| c | c | c | c | c | c |}
                \hline
                \textbf{} & \textbf{Sensor Type} & \textbf{Image Dimensions} & \textbf{Set A (d x w)} & \textbf{Set B (d x w)} & \textbf{Resolution}\\ \hline
                \textbf{DB1} & Optical Sensor & 640x480 (307 Kpixels) & 8x100 & 8x10 & 500 dpi \\ \hline
                \textbf{DB2} & Optical Sensor & 328x364 (119 Kpixels) & 8x100 & 8x10 & 500 Kdpi \\\hline
                \textbf{DB3} & Thermal sweeping Sensor & 300x480 (144 Kpixels) & 8x100 & 8x10 & 512 dpi \\\hline
                \textbf{DB4} & SFinGe v3.0 & 288x384 (108 Kpixels) & 8x100 & 8x10 & about 500 dpi \\ \hline
                \end{tabular}
            }
        \end{table}

        \begin{table}[ht!]
        \centering
            \caption{FVC2006 dataset}\label{t5}
            \resizebox{\textwidth}{!}{
                \begin{tabular}{| c | c | c | c | c | c |}
                \hline
                \textbf{} & \textbf{Sensor Type} & \textbf{Image Dimensions} & \textbf{Set A (d x w)} & \textbf{Set B (d x w)} & \textbf{Resolution}\\ \hline
                \textbf{DB1} & Electric Field sensor & 96x96 (9 Kpixels) & 140x12 & 10x12 & 250 dpi \\ \hline
                \textbf{DB2} & Optical Sensor & 400x560 (224 Kpixels)	& 140x12 & 10x12 & 569 dpi \\\hline
                \textbf{DB3} & Thermal sweeping Sensor & 400x500 (200 Kpixels) & 140x12 & 10x12 & 500 dpi \\\hline
                \textbf{DB4} & SFinGe v3.0 & 288x384 (108 Kpixels) & 140x12 & 10x12 & about 500 dpi \\ \hline
                \end{tabular}
            }
        \end{table}

        While processing the FVC2006 DB1\_A, we have resized all the images in dataset from 96 x 96 to 400 x 400 since due to small size of the images, we were unable to extract minutiae points from them.

\subsection{Experimental Setup}
We have used Python to test the performance and the time of the scheme in the MacOS environment. The MAC used in the experiment is configured as Apple M1 Pro, and the memory is 16 GB and 6400MHz LPDDR5 with 8-core CPU and 14-core GPU. The size of the proof is found to be around 200 bytes and the verifier time is around $3 ms$ using this setup.

\subsection{Evaluation Metrics}
Based on our experiments and analysis, we observed that zkSNARK provides a promising solution for secure and privacy-preserving fingerprint authentication. To prove the efficacy of our proposed approach, we have evaluated various performance metrics.

\subsubsection{Accuracy vs Threshold}
\begin{figure}[!htbp]
    \centering
    \subfigure[]{\includegraphics[width=0.49\textwidth]{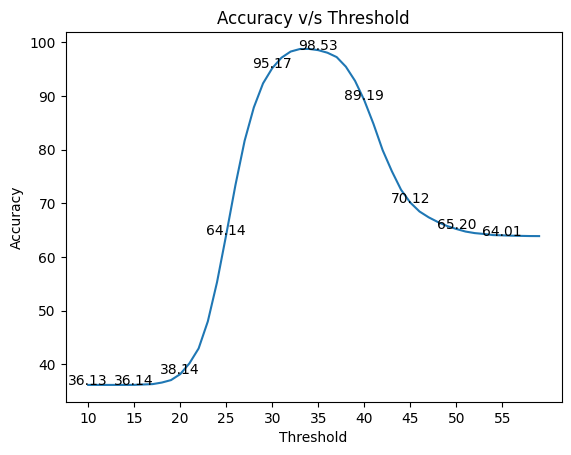}}
    \subfigure[]{\includegraphics[width=0.49\textwidth]{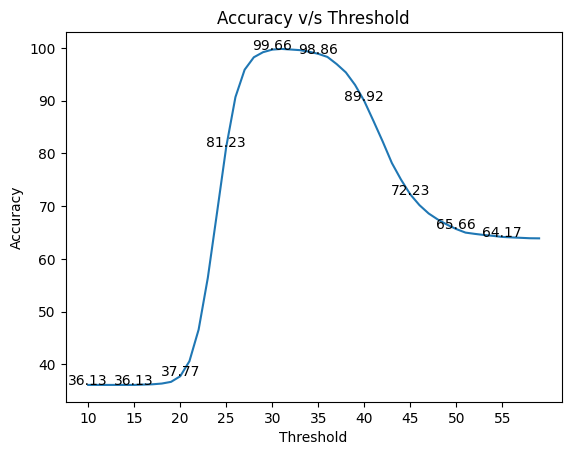}}
    \subfigure[]{\includegraphics[width=0.49\textwidth]{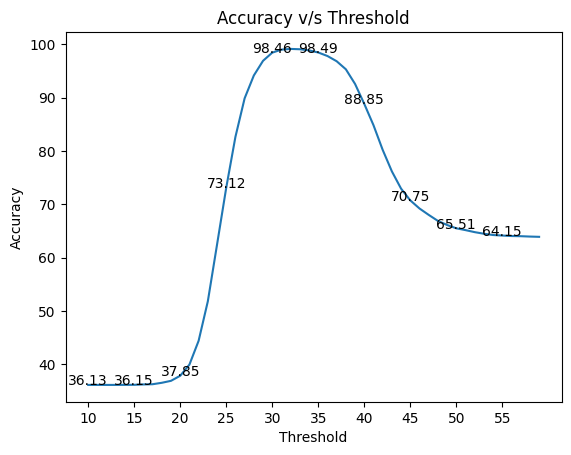}}
    \subfigure[]{\includegraphics[width=0.49\textwidth]{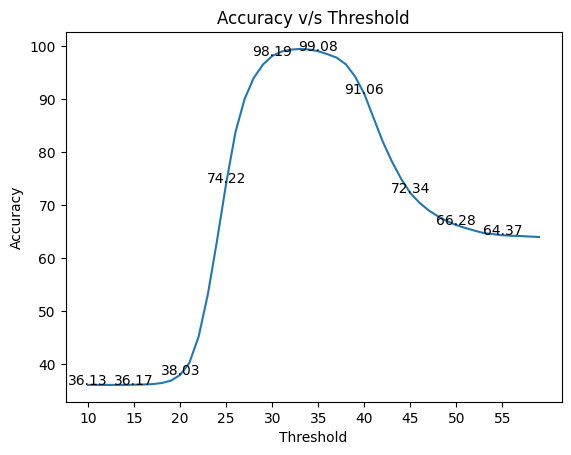}}
    \caption{Accuracy vs Threshold of FVC2002 (a) DB1\_A (b) DB2\_A (c) DB3\_A (d) DB4\_A}
    \label{f5}
\end{figure}

\begin{figure}[!htbp]
    \centering
    \subfigure[]{\includegraphics[width=0.49\textwidth]{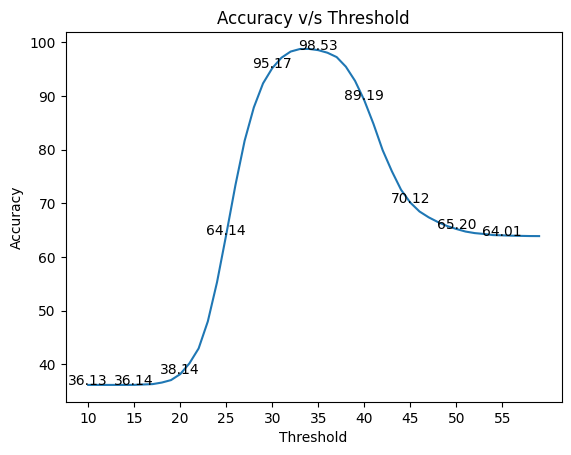}}
    \subfigure[]{\includegraphics[width=0.49\textwidth]{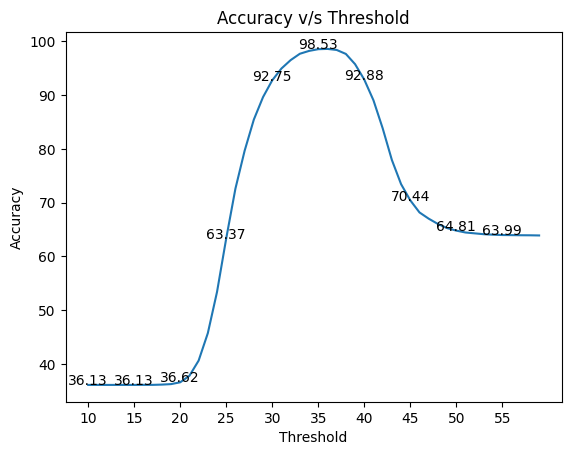}}
    \subfigure[]{\includegraphics[width=0.49\textwidth]{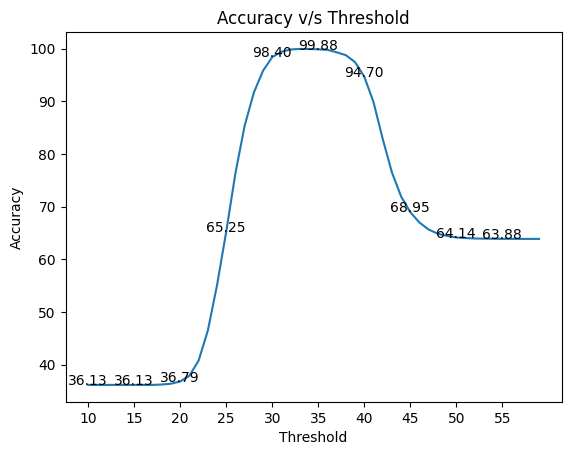}}
    \subfigure[]{\includegraphics[width=0.49\textwidth]{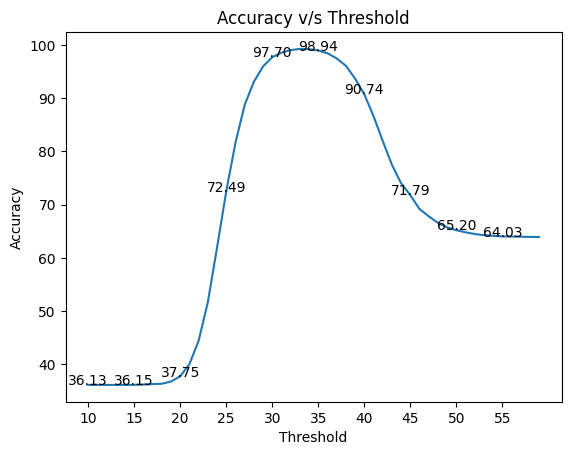}}
    \caption{Accuracy vs Threshold of FVC2004 (a) DB1\_A (b) DB2\_A (c) DB3\_A (d) DB4\_A}
    \label{f6}
\end{figure}

\begin{figure}[!htbp]
    \centering
    \subfigure[]{\includegraphics[width=0.49\textwidth]{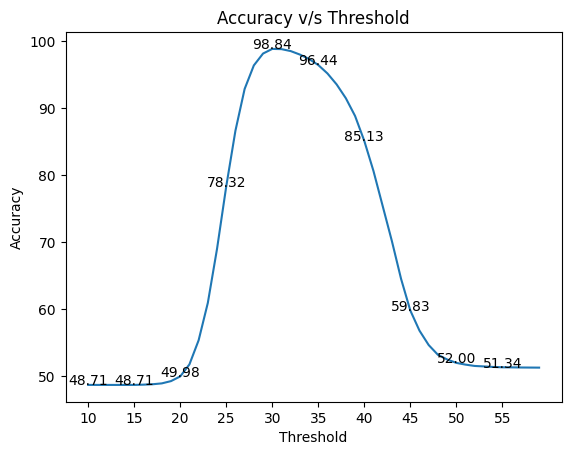}}
    \subfigure[]{\includegraphics[width=0.49\textwidth]{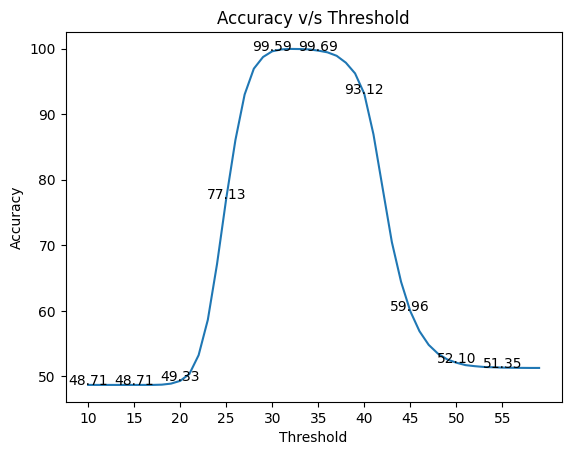}}
    \subfigure[]{\includegraphics[width=0.49\textwidth]{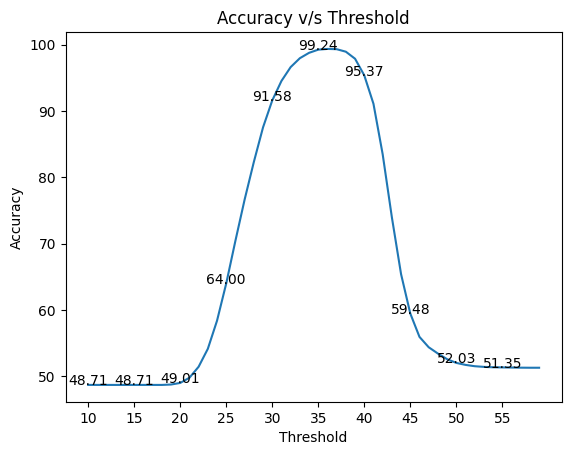}}
    \subfigure[]{\includegraphics[width=0.49\textwidth]{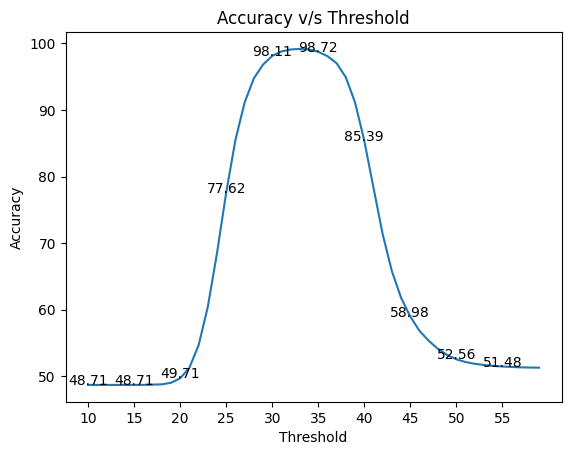}}
    \caption{Accuracy vs Threshold of FVC2006 (a) DB1\_A (b) DB2\_A (c) DB3\_A (d) DB4\_A}
    \label{f7}
\end{figure}

In our K-NNS-based approach, we produce a Global Matching score(GMS) and if that GMS is greater than some threshold only then we validate it as a successful authentication. We varied the threshold from 10\% to 60\% and computed the accuracy of the model for the corresponding threshold as can be seen across all the four datasets of FVC2002 in the figure \ref{f5}, of FVC2004 in the figure \ref{f6} and of FVC2006 in the figure \ref{f7}. The figures display variation of accuracy with the size of threshold for all the databases.\par
Based on the findings we selected 30\% to be the threshold value resulting in an average accuracy of 99.01\% over the FVC2002 dataset and 35\% to be the threshold value in an average accuracy of 98.97\% and 98.52\% over FVC2004 and FVC2006 respectively.

\subsubsection{False Acceptance Rate \& Genuine Acceptance Rate vs Threshold}
\begin{figure}[!htbp]
    \centering
    \subfigure[]{\includegraphics[width=0.49\textwidth]{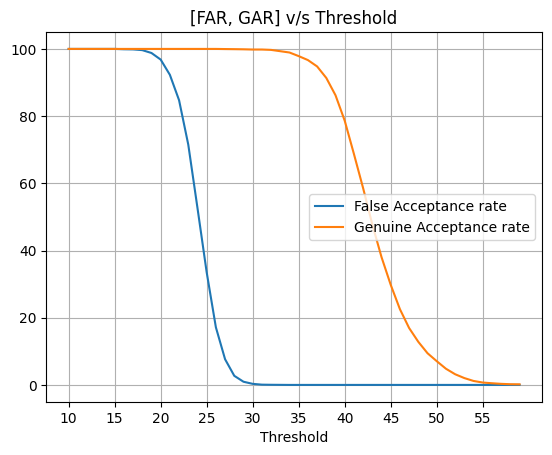}}
    \subfigure[]{\includegraphics[width=0.49\textwidth]{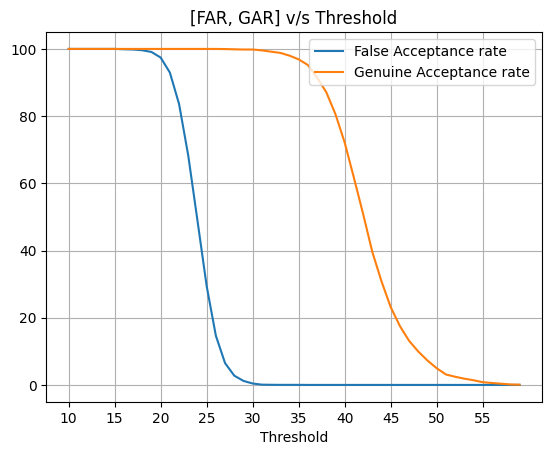}}
    \subfigure[]{\includegraphics[width=0.49\textwidth]{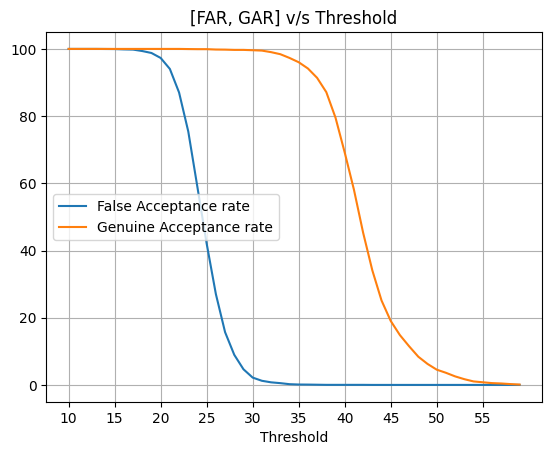}}
    \subfigure[]{\includegraphics[width=0.49\textwidth]{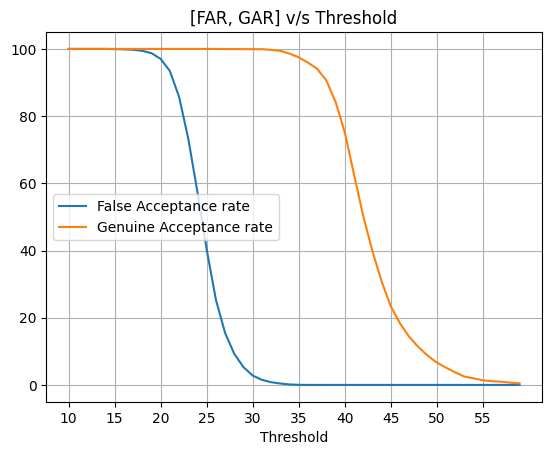}}
    \caption{Genuine Acceptance Rate and False Acceptance Rate vs Threshold of FVC2002 (a) DB1\_A (b) DB2\_A (c) DB3\_A (d) DB4\_A}
    \label{f8}
\end{figure}

\begin{figure}[!htbp]
    \centering
    \subfigure[]{\includegraphics[width=0.49\textwidth]{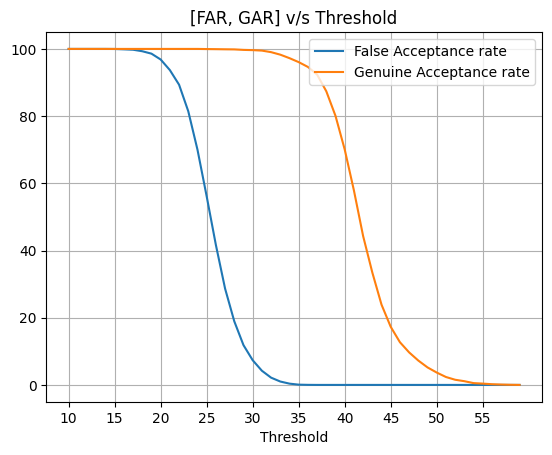}}
    \subfigure[]{\includegraphics[width=0.49\textwidth]{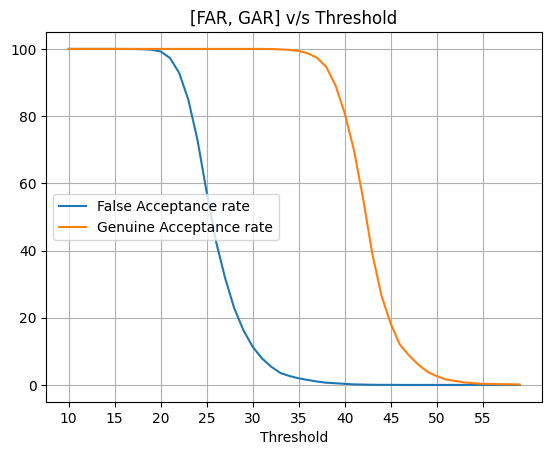}}
    \subfigure[]{\includegraphics[width=0.49\textwidth]{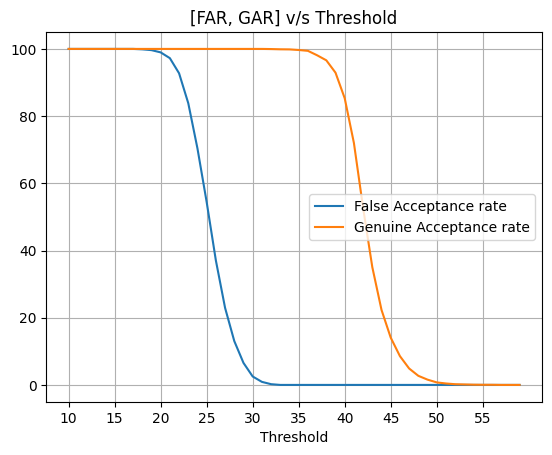}}
    \subfigure[]{\includegraphics[width=0.49\textwidth]{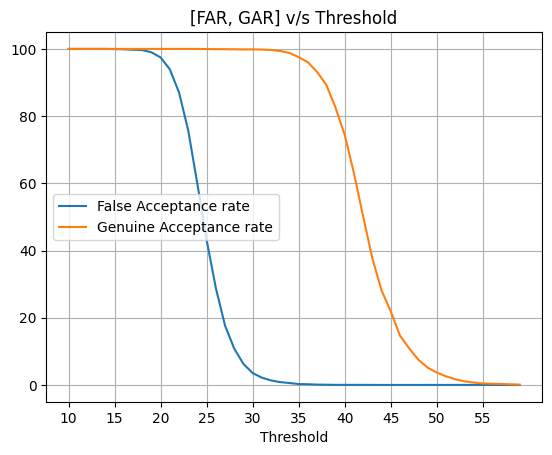}}
    \caption{Genuine Acceptance Rate and False Acceptance Rate vs Threshold of FVC2004 (a) DB1\_A (b) DB2\_A (c) DB3\_A (d) DB4\_A}
    \label{f9}
\end{figure}

\begin{figure}[!htbp]
    \centering
    \subfigure[]{\includegraphics[width=0.49\textwidth]{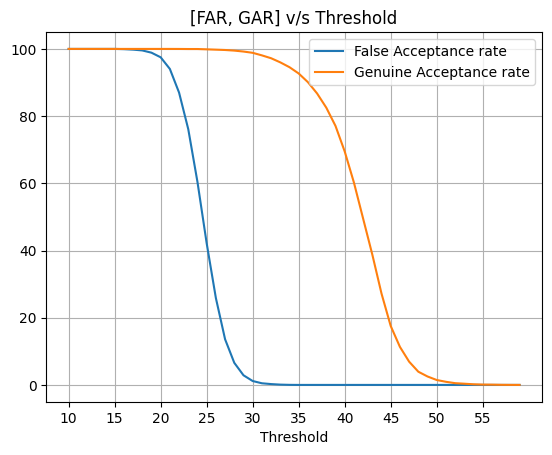}}
    \subfigure[]{\includegraphics[width=0.49\textwidth]{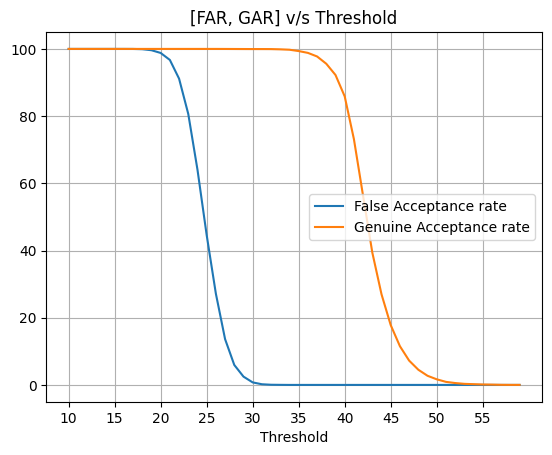}}
    \subfigure[]{\includegraphics[width=0.49\textwidth]{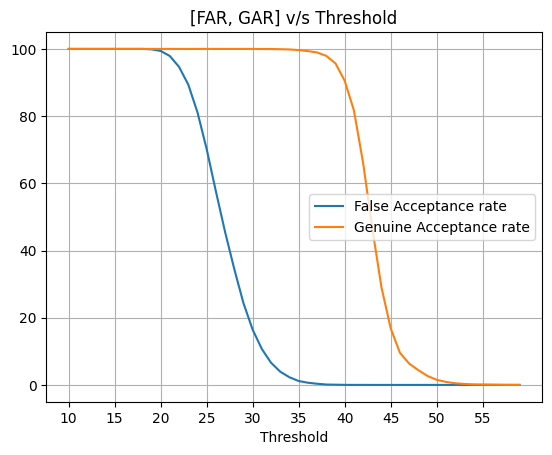}}
    \subfigure[]{\includegraphics[width=0.49\textwidth]{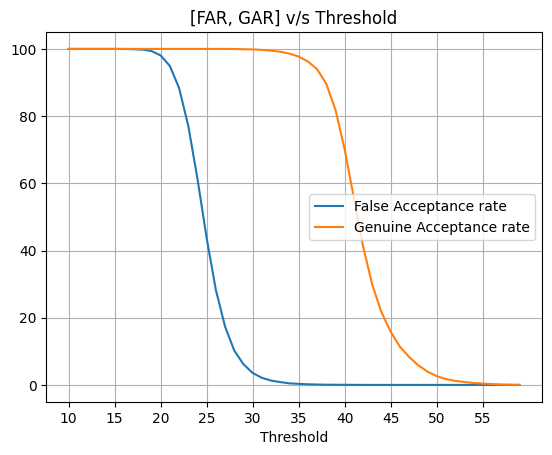}}
    \caption{Genuine Acceptance Rate and False Acceptance Rate vs Threshold of FVC2006 (a) DB1\_A (b) DB2\_A (c) DB3\_A (d) DB4\_A}
    \label{f10}
\end{figure}

In the case of a biometric authentication system, the false acceptance rate must be close to 0 so that no false user is authenticated as true. Thus, we have plotted the genuine acceptance rate and false acceptance rate with the corresponding threshold four datasets of FVC2002 in the figure \ref{f8}, of FVC2004 in the figure \ref{f9} and of FVC2006 in the figure \ref{f10}. From the graph, we can select the threshold with the least false acceptance rate and a good true acceptance rate.

\subsubsection{Receiver Operating Characteristic curve}
\begin{figure}[!htbp]
    \centering
    \subfigure[]{\includegraphics[width=0.49\textwidth]{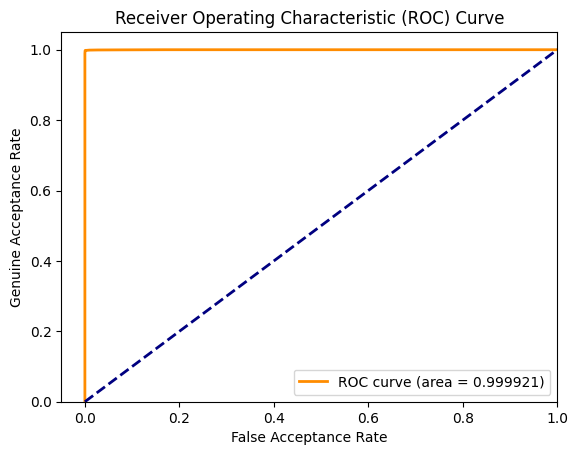}}
    \subfigure[]{\includegraphics[width=0.49\textwidth]{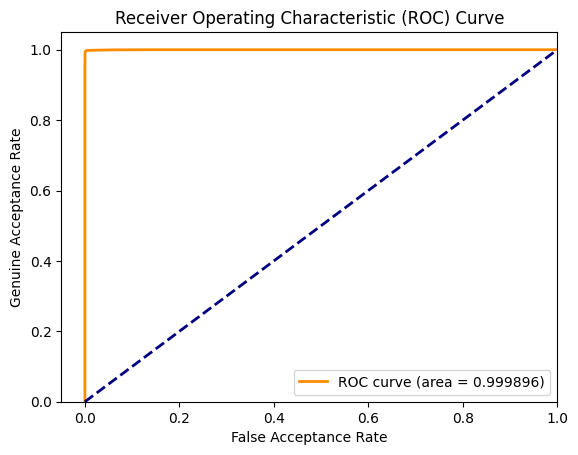}}
    \subfigure[]{\includegraphics[width=0.49\textwidth]{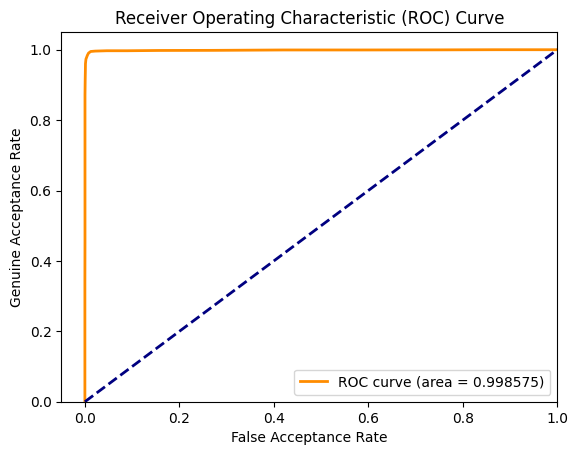}}
    \subfigure[]{\includegraphics[width=0.49\textwidth]{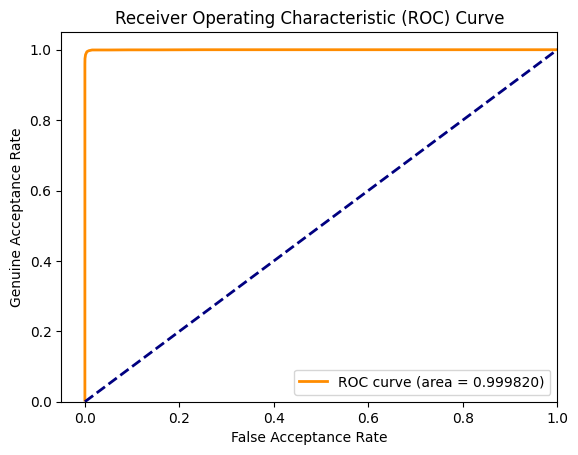}}
    \caption{Receiver Operating Characteristic curve of FVC2002 (a) DB1\_A (b) DB2\_A (c) DB3\_A (d) DB4\_A}
    \label{f11}
\end{figure}

\begin{figure}[!htbp]
    \centering
    \subfigure[]{\includegraphics[width=0.49\textwidth]{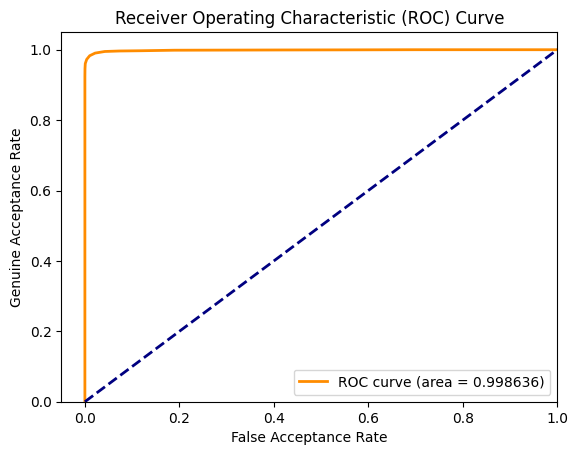}}
    \subfigure[]{\includegraphics[width=0.49\textwidth]{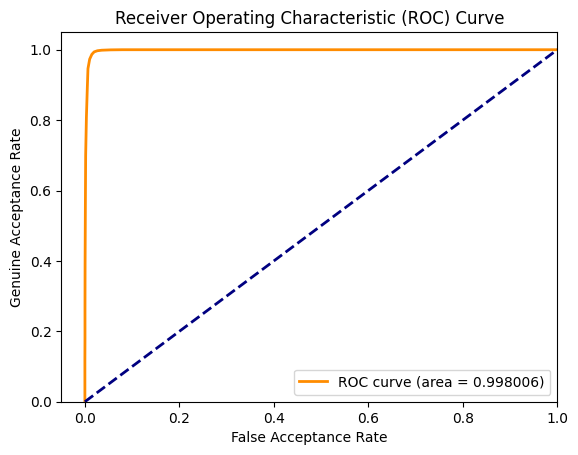}}
    \subfigure[]{\includegraphics[width=0.49\textwidth]{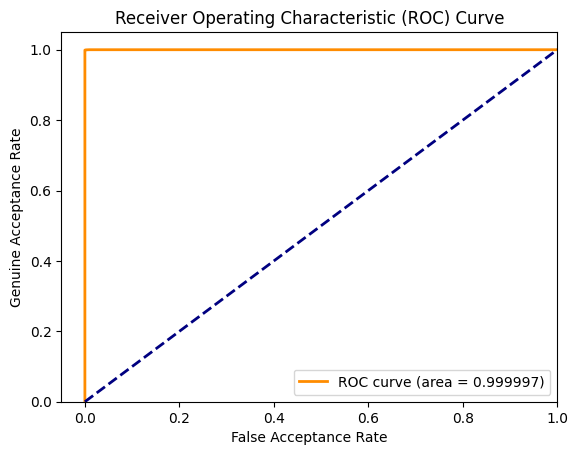}}
    \subfigure[]{\includegraphics[width=0.49\textwidth]{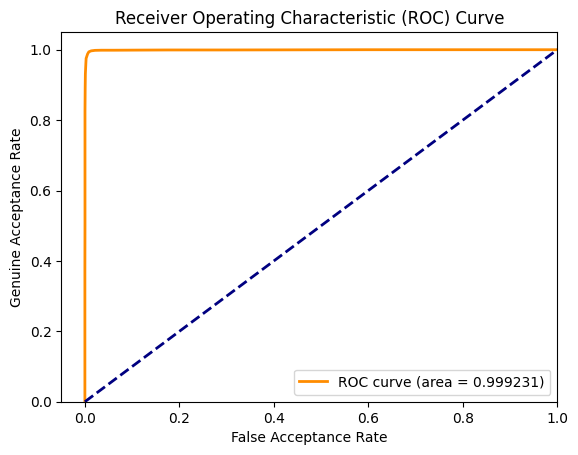}}
    \caption{Receiver Operating Characteristic curve of FVC2004 (a) DB1\_A (b) DB2\_A (c) DB3\_A (d) DB4\_A}
    \label{f12}
\end{figure}

\begin{figure}[!htbp]
    \centering
    \subfigure[]{\includegraphics[width=0.49\textwidth]{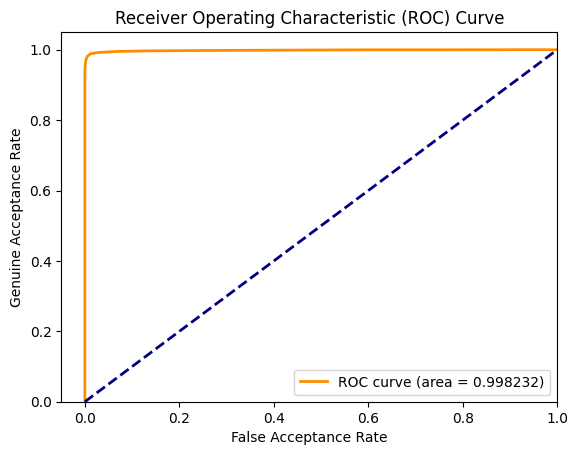}}
    \subfigure[]{\includegraphics[width=0.49\textwidth]{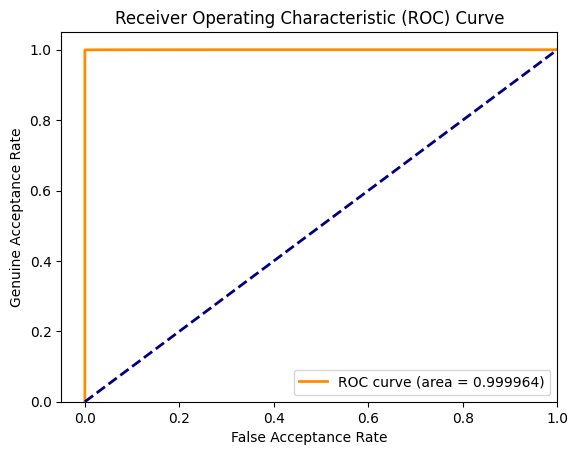}}
    \subfigure[]{\includegraphics[width=0.49\textwidth]{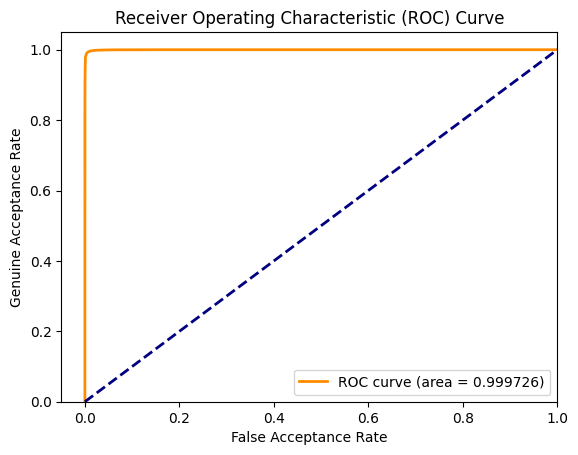}}
    \subfigure[]{\includegraphics[width=0.49\textwidth]{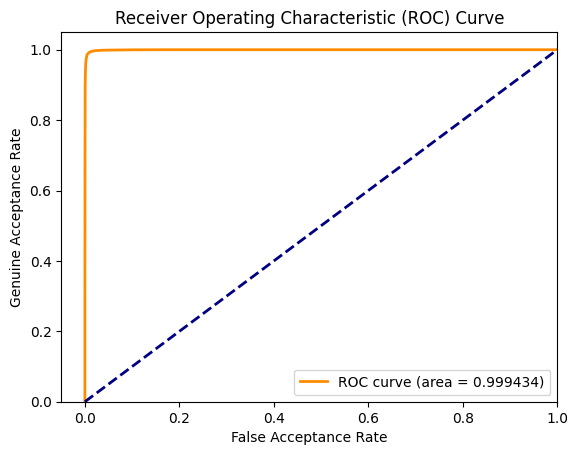}}
    \caption{Receiver Operating Characteristic curve of FVC2006 (a) DB1\_A (b) DB2\_A (c) DB3\_A (d) DB4\_A}
    \label{f13}
\end{figure}

The Receiver Operating Characteristic (ROC) curve plots the true acceptance rate (TAR) against the false acceptance rate (FAR) for different threshold values. The TAR is the proportion of genuine fingerprints that are correctly identified as such, while the FPR is the proportion of impostor fingerprints that are incorrectly identified as genuine. We have plotted the ROC curves as shown in figure \ref{f11}, \ref{f12} and \ref{f13} for FVC2002, FVC2004 and FVC2006 respectively. The ROC curve can be used to determine the optimal threshold value that balances the trade-off between TPR and FPR and to compare the performance of different fingerprint authentication systems.\\

\subsubsection{zk-SNARK Proof}
The size of the proof and the verifier time are two important metrics for evaluating the performance using zk-SNARK. The size of the proof is the amount of data that the prover needs to send to the verifier in order to prove their identity. A smaller proof size is desirable because it reduces communication overhead and makes the authentication process more efficient. We have observed the proof size to be around 200 bytes. The verifier time is the amount of time it takes for the verifier to verify the proof. A shorter verifier time is desirable because it makes the authentication process more responsive. We have observed the verifier time to be around $3ms$ verifier time.

\section{Future Scope and Conclusion}
We have proposed a novel way to incorporate zk-SNARK for Blockchain-based identity management systems using off-chain computations. We use a KNNS-based approach on all the datasets of the FVC2002, FVC2004 and FVC2006 to generate a cancelable template for secure, faster and robust biometric (fingerprint) registration and authentication and have obtained an average accuracy of 99.01\% over the FVC2002 dataset, an average accuracy of 98.97\% over FVC2004 and  an average accuracy of 98.52\%  over FVC2006 for fingerprint authentication. We have also observed the proof size to be around 200 bytes and the verifier time to be around $3ms$.

This work is not exhaustive, as the fields of biometric authentication and zero-knowledge proofs are continuously evolving. While fingerprint authentication is a powerful authentication factor, it may not be sufficient on its own. More work could be done towards a multi-modal identity management system by exploring ways to integrate fingerprint authentication with other authentication factors, such as facial recognition or iris scanning, creating a more robust authentication system.

As a future scope, we wish to incorporate the model with some existing centrally administered identity management systems. Such models can also be tested and optimize the effectiveness of the approach on alternative datasets, such as Sokoto Coventry Fingerprint \cite{shehu2018sokoto}, for accurate and faster authentication. Furthermore, different zk-snark algorithms, such as Sonic, Halo, and Plonk, or building a customized version could also be explored.

The project can also be used to explore various new use cases for fingerprint authentication using blockchain and zkSNARK. For example, the technology can be used in identity verification for online payments, voting systems, or other applications where security and privacy are critical. Financial institutions can make use of this research work to streamline their KYC (Know Your Customer) processes, reducing the time and resources required for customer onboarding and improving the overall customer experience.

 \bibliographystyle{elsarticle-num} 
 \bibliography{cas-refs}





% \end{thebibliography}
\end{document}